\documentclass[superscriptaddress,preprintnumbers]{revtex4}

\usepackage{graphicx,epsfig}
\def\beq{\begin{equation}}
\def\eeq{\end{equation}}
\def\beqar{\begin{eqnarray}}
\def\eeqar{\end{eqnarray}}
\def\barr#1{\begin{array}{#1}}
\def\earr{\end{array}}
\def\bfi{\begin{figure}}
\def\efi{\end{figure}}
\def\btab{\begin{table}}
\def\etab{\end{table}}
\def\bce{\begin{center}}
\def\ece{\end{center}}

\def\text{\textstyle}

\def\arraystretch{1.4}

\def\be{\beta}

\def\refeq#1{\mbox{eq.~(\ref{#1})}}
\def\refeqs#1{\mbox{eqs.~(\ref{#1})}}

\def\reffis#1{\mbox{Figs.~\ref{#1}}}

\def\citere#1{\mbox{Ref.~\cite{#1}}}
\def\citeres#1{\mbox{Refs.~\cite{#1}}}

\newcommand{\GeV}{\unskip\,\mathrm{GeV}}

\newcommand{\TeV}{\unskip\,\mathrm{TeV}}

\def\mathswitchr#1{\relax\ifmmode{\mathrm{#1}}\else$\mathrm{#1}$\fi}

\newcommand{\PA}{\mathswitchr A}

\newcommand{\Ph}{\mathswitchr h}

\newcommand{\Pt}{\mathswitchr t}

\def\mathswitch#1{\relax\ifmmode#1\else$#1$\fi}

\newcommand{\Mt}{\mathswitch {m_\Pt}}

\newcommand{\mh}{\mathswitch {m_\Ph}}

\newcommand{\MA}{\mathswitch {M_\PA}}

\newcommand{\mt}{\Mt}

\newcommand{\tsf}{\theta\kern-.20em_{\tilde{f}}}
\newcommand{\tsfp}{\theta\kern-.20em_{\tilde{f}\prime}}
\newcommand{\tsq}{\theta\kern-.15em_{\tilde{q}}}

\newcommand{\lsim}
{\;\raisebox{-.3em}{$\stackrel{\displaystyle <}{\sim}$}\;}
\newcommand{\gsim}
{\;\raisebox{-.3em}{$\stackrel{\displaystyle >}{\sim}$}\;}

\newcommand{\fh}{\textsl{FeynHiggs}}
\newcommand{\subh}{\textsl{subhpole}}

\newcommand{\cp}{{\cal CP}}

\newcommand{\VL}{\left( \begin{array}{c}}
\newcommand{\VR}{\end{array} \right)}
\newcommand{\ML}{\left( \begin{array}{cc}}
\newcommand{\MLd}{\left( \begin{array}{ccc}}
\newcommand{\MLv}{\left( \begin{array}{cccc}}
\newcommand{\MR}{\end{array} \right)}

\newcommand{\BC}{\begin{center}}
\newcommand{\EC}{\end{center}}
\newcommand{\BE}{\begin{equation}}
\newcommand{\EE}{\end{equation}}
\newcommand{\BEA}{\begin{eqnarray}}
\newcommand{\BEAnn}{\begin{eqnarray*}}
\newcommand{\EEA}{\end{eqnarray}}
\newcommand{\EEAnn}{\end{eqnarray*}}
\newcommand{\non}{\nonumber}
\newcommand{\id}{{\rm 1\kern-.12em
\rule{0.3pt}{1.5ex}\raisebox{0.0ex}{\rule{0.1em}{0.3pt}}}}

\def\draftdate{\relax}
\def\mda{\relax}
\def\mua{\relax}
\def\mla{\relax}
\def\draft{
\def\thtystars{******************************}
\def\sixtystars{\thtystars\thtystars}
\typeout{}
\typeout{\sixtystars**}
\typeout{* Draft mode!
         For final version remove \protect\draft\space in source file
*}
\typeout{\sixtystars**}
\typeout{}
\def\draftdate{\today}
\def\mua{\marginpar[\boldmath\hfil$\uparrow$]%
                   {\boldmath$\uparrow$\hfil}%
                    \typeout{marginpar: $\uparrow$}\ignorespaces}
\def\mda{\marginpar[\boldmath\hfil$\downarrow$]%
                   {\boldmath$\downarrow$\hfil}%
                    \typeout{marginpar: $\downarrow$}\ignorespaces}
\def\mla{\marginpar[\boldmath\hfil$\rightarrow$]%
                   {\boldmath$\leftarrow $\hfil}%
                    \typeout{marginpar:
$\leftrightarrow$}\ignorespaces}
\def\Mua{\marginpar[\boldmath\hfil$\Uparrow$]%
                   {\boldmath$\Uparrow$\hfil}%
                    \typeout{marginpar: $\Uparrow$}\ignorespaces}
\def\Mda{\marginpar[\boldmath\hfil$\Downarrow$]%
                   {\boldmath$\Downarrow$\hfil}%
                    \typeout{marginpar: $\Downarrow$}\ignorespaces}
\def\Mla{\marginpar[\boldmath\hfil$\Rightarrow$]%
                   {\boldmath$\Leftarrow $\hfil}%
                    \typeout{marginpar:
$\Leftrightarrow$}\ignorespaces}
\overfullrule 5pt
\oddsidemargin -15mm
\marginparwidth 29mm
}

\begin{document}

\begin{flushright}
BNL--HET--02/6, CERN--TH/2002--020\\
DCPT/02/16, DESY 02--022\\
FERMILAB-Conf-02/011-T, HEPHY-PUB 751\\
IPPP/02/08, PM/01--69, SLAC-PUB-9134\\
UCD-2002-01, UFIFT-HEP-02-2\\
UMN--TH--2043/02, ZU-TH 3/02
\end{flushright}


\bibliographystyle{revtex}

%

\title{The Snowmass Points and Slopes: Benchmarks for SUSY Searches}



\author{B.C.~Allanach}
\affiliation{CERN, Geneva, Switzerland}

\author{M.~Battaglia}
\affiliation{CERN, Geneva, Switzerland}   

\author{G.A.~Blair}
\affiliation{Royal Holloway, Univ.\ of London, UK}

\author{M.~Carena}
\affiliation{Fermilab, Batavia IL, U.S.A.}    

\author{A.~De~Roeck}
\affiliation{CERN, Geneva, Switzerland}  

\author{A.~Dedes}
\affiliation{University of Bonn, Germany}

\author{A.~Djouadi}
\affiliation{LPMT, Universit\'e de Montpellier II, 
France}


\author{D.~Gerdes}
\affiliation{Dept.\ of Physics, Univ.~of Michigan, U.S.A.} 

\author{N.~Ghodbane}
\affiliation{DESY, Hamburg, Germany} 



\author{J.~Gunion}
\affiliation{Department of Physics, University of California at Davis,
Davis CA, U.S.A.}

\author{H.E.~Haber}
\affiliation{Santa Cruz Inst.~for Part.~Phys., UCSC, Santa Cruz CA, U.S.A.} 

\author{T.~Han}
\affiliation{Dept.\ of Physics, Univ.~of Wisconsin, Madison, U.S.A.}  

\author{S.~Heinemeyer}
\affiliation{HET Physics Dept., Brookhaven Natl.\ Lab., NY, U.S.A.}

\author{J.L.~Hewett}
\affiliation{Stanford Linear Accelerator Center, Stanford University, 
Stanford CA, U.S.A.}

\author{I.~Hinchliffe}
\affiliation{Lawrence Berkeley Natl.\ Lab., Berkeley CA, U.S.A.}

\author{J.~Kalinowski}
\affiliation{Institute of Theoretical Physics, UW, Warsaw, Poland} 


\author{H.E.~Logan}
\affiliation{Fermilab, Batavia IL, U.S.A.} 

\author{S.P.~Martin}
\affiliation{Fermilab, Batavia IL, U.S.A.}
\affiliation{Northern Illinois University, DeKalb IL, U.S.A}

\author{H.-U.~Martyn}
\affiliation{I.~Physikalisches Institut, RWTH Aachen, Germany} 

\author{K.T.~Matchev}
\affiliation{CERN, Geneva, Switzerland}
\affiliation{Department of Physics, University of Florida,
             Gainesville FL, U.S.A.}

\author{S.~Moretti}
\affiliation{CERN, Geneva, Switzerland}   
\affiliation{Institute for Particle Physics Phenomenology, Durham, UK}

\author{F.~Moortgat}
\affiliation{University of Antwerpen, Wilrijk, Belgium}

\author{G.~Moortgat-Pick}
\affiliation{DESY, Hamburg, Germany}

\author{S.~Mrenna}
\affiliation{Fermilab, Batavia IL, U.S.A.}

\author{U.~Nauenberg}
\affiliation{Physics Dept., University of Colorado, Boulder CO, U.S.A.}    

\author{Y.~Okada}
\affiliation{KEK, Tsukuba, Ibaraki, Japan}

\author{K.A.~Olive}
\affiliation{University of Minnesota, Minneapolis MN, U.S.A.}

\author{W.~Porod}
\affiliation{Inst.~f.~Hochenergiephysik, \"Oster.~Akademie d.~Wissenschaften,
      Vienna, Austria}
\affiliation{Inst.~f\"ur Theor.~Physik, Universit\"at Z\"urich,
Switzerland}

\author{M.~Schmitt}
\affiliation{Northwestern University, Evanston IL, U.S.A.}

\author{S.~Su}
\affiliation{California Institute of Technology, Pasadena CA, U.S.A.}


\author{C.E.M.~Wagner}
\affiliation{High Energy Division, Argonne Natl.~Lab., Argonne IL, U.S.A.}
\affiliation{Enrico Fermi Institute, Univ.\ of Chicago, Chicago IL, U.S.A.}

\author{G.~Weiglein}
\email[]{Georg.Weiglein@durham.ac.uk}
\affiliation{Institute for Particle Physics Phenomenology, Durham, UK}

\author{J.~Wells}
\affiliation{Department of Physics, University of California at Davis,
Davis CA, U.S.A.}

\author{G.W.~Wilson}
\affiliation{University of Kansas, Lawrence KS, U.S.A.}

\author{P.~Zerwas}
\affiliation{DESY, Hamburg, Germany}


\date{February 22, 2002}

\begin{abstract}
The ``Snowmass Points and Slopes'' (SPS) are a set of benchmark points
and parameter lines in the MSSM parameter space 
corresponding to different scenarios in the search
for Supersymmetry at present and future experiments. This set of
benchmarks was agreed upon at the 2001 ``Snowmass Workshop on the Future of
Particle Physics'' as a consensus based on different existing proposals.
\end{abstract}

\maketitle



\section{Why benchmarks --- which benchmarks?}

In the unconstrained version of the Minimal Supersymmetric extension of
the Standard Model (MSSM) no particular Supersymmetry (SUSY) breaking 
mechanism is assumed, but rather a parameterization of all possible soft 
SUSY breaking terms is used.
This leads to more than a hundred parameters
(masses, mixing angles, phases) in this model in addition to the ones of
the Standard Model. The currently most popular SUSY breaking
mechanisms are minimal supergravity (mSUGRA)~\cite{msugra},
gauge-mediated SUSY breaking (GMSB)~\cite{gmsb}, and anomaly-mediated
SUSY breaking (AMSB)~\cite{amsb}. 
In these scenarios SUSY breaking happens in a hidden
sector and is mediated to the visible sector (i.e.\ the MSSM) in
different ways: via gravitational interactions in the mSUGRA scenario,
via gauge interactions in the GMSB scenario, and via the super-Weyl
anomaly in the AMSB scenario. Assuming one of these SUSY breaking
mechanisms leads to a drastic reduction of the number of 
parameters compared to the MSSM case. The mSUGRA scenario is 
characterized by four parameters
and a sign, the scalar mass parameter $m_0$, the gaugino mass parameter 
$m_{1/2}$, the trilinear coupling $A_0$, the ratio of the Higgs vacuum
expectation values, $\tan\beta$, and the sign of the supersymmetric Higgs 
mass parameter, $\mu$. The parameters of the (minimal) GMSB scenario are 
the messenger mass $M_{\rm mes}$, the messenger index $N_{\rm mes}$,
the universal soft SUSY breaking mass scale felt by the low-energy
sector, $\Lambda$, as well as $\tan\beta$ and $\mathrm{sign}(\mu)$.
The (minimal) AMSB scenario has the parameters $m_{\rm aux}$, which sets
the overall scale of the SUSY particle masses (given by the vacuum
expectation value of the auxiliary field in the supergravity multiplet),
$\tan\beta$, $\mathrm{sign}(\mu)$, and $m_0$, where the latter is a
phenomenological parameter introduced in order to keep the squares
of slepton masses positive. The mass spectra of the SUSY particles 
in these scenarios are obtained via renormalization
group running from the scale of the high-energy parameters of the
SUSY-breaking scenario to the weak scale.
The low-energy parameters obtained in this way are then used as input 
for calculating the
predictions for the production cross sections and for the decay
branching ratios of the SUSY particles.

While a detailed scanning over the more-than-hundred-dimensional parameter 
space of the MSSM is clearly not practicable, even a sampling of the three- 
or four-dimensional parameter space of the above-mentioned SUSY breaking
scenarios is beyond the present capabilities for phenomenological
studies, in particular when it comes to simulating
experimental signatures within the detectors. For this reason one often
resorts to specific benchmark scenarios, i.e.\ one studies only specific
parameter points or at best samples a one-dimensional parameter space
(the latter is sometimes called a model line~\cite{modelline}), which
exhibit specific characteristics of the MSSM parameter space.
Benchmark scenarios of this kind are often used, for instance, for studying 
the performance of different experiments at the same collider. Similarly,
detailed experimental simulations of sparticle production with identical
MSSM parameters in the
framework of different colliders can be very helpful for developing
strategies for combining pieces of information obtained at 
different machines.

The question of which parameter choices are useful as benchmark
scenarios depends on the purpose of the actual investigation. If one is
interested, for instance, in setting exclusion limits on the SUSY
parameter space from the non-observation of SUSY signals at the
experiments performed up to now, it is useful to use a benchmark
scenario which gives rise to ``conservative'' exclusion bounds. An
example of a benchmark scenario of this kind is the
$\mh^{\mathrm{max}}$-scenario~\cite{lephiggsbenchmarks} used for the
Higgs search at 
LEP~\cite{lepbench} and the Tevatron~\cite{tevbench}. It gives rise to 
maximal values of the lightest
$\cp$-even Higgs-boson mass (for fixed values of the top-quark mass and
the SUSY scale) and thus allows one to set conservative bounds on
$\tan\beta$ and $\MA$ (the mass of the $\cp$-odd Higgs boson)~\cite{tbexcl}. 
Another application of benchmark scenarios is to study ``typical''
experimental signatures of SUSY models and to investigate the
experimental sensitivities and the achievable experimental precisions
for these cases. For this purpose it seems reasonable to choose
``typical'' (a notion which is of course difficult to define) and
theoretically well motivated parameters of 
certain SUSY-breaking scenarios. Examples of this kind are the benchmark
scenarios used so far for investigating SUSY searches at the
LHC~\cite{benchsnow96,tdratlcms}, the Tevatron~\cite{tevsugra}
and at a future Linear Collider~\cite{teslatdr}.
As a further possible goal of benchmark scenarios, one can choose them
so that they account for a wide variety of SUSY phenomenology. 
For this purpose, one could for instance analyse  
SUSY with R-parity breaking, investigate effects of non-vanishing $\cp$~phases,
or inspect non-minimal SUSY models. In this context it can also be
useful to consider ``pathological'' regions of parameter space or
``worst-case'' scenarios. Examples for this are the ``large-$|\mu|$
scenario'' for the Higgs search at LEP~\cite{lephiggsbenchmarks} and the 
Tevatron~\cite{hadhiggsbenchmarks}, for which the decay $h \to b \bar b$ 
can be significantly suppressed, 
or a scenario where the Higgs boson has a large branching fraction into
invisible decay modes at the LHC (see e.g.\ \citere{fawzi}).

A related issue concerning the definition of appropriate benchmarks is
whether a benchmark scenario chosen for investigating physics at a
certain experiment or for testing a certain sector of the theory should be 
compatible with additional information from other experiments (or
concerning other sectors of the theory). This refers in
particular to constraints from cosmology (by demanding that SUSY should
give rise to an acceptable dark matter density~\cite{cdm}) and low-energy
measurements such as the rate for $b \to s \gamma$~\cite{bsg} and the 
anomalous magnetic moment of the muon, $g_{\mu} -2$~\cite{gminus2} (see
\citere{gminus2th} for the updated SM prediction for $g_{\mu} -2$). On
the one hand, applying 
constraints of this kind gives rise to ``more realistic'' benchmark
scenarios. On the other hand, one relies in this way on further
assumptions (and has to take account of experimental and theoretical
uncertainties related to these additional constraints), and it could
eventually turn out that one has inappropriately narrowed down the range of
possibilities by applying these constraints. This applies in
particular if slight modifications of the SUSY breaking scenarios are
allowed that have a minor impact on collider phenomenology but could
significantly alter the bounds from cosmology and low-energy
experiments. For instance, the presence of small flavor mixing terms in the
SUSY Lagrangian could severely affect the prediction for BR($b \to s \gamma$),
while allowing a small amount of R-parity violation in the model would
strongly affect the constraints from dark matter relic abundance
while leaving collider phenomenology essentially 
unchanged. In the context of additional constraints one also has to
decide on the level of fine-tuning of parameters (as a measure to 
distinguish between ``more natural'' and ``less natural'' parameter 
choices) one should tolerate in a benchmark scenario.

The extent to which additional constraints of this kind should
be applied to possible benchmark scenarios is related to the actual purpose
of the benchmark scenario. For setting exclusion bounds in a particular
sector (e.g.\ the Higgs sector) it seems preferable to apply constraints
only from this sector. Similarly, relaxing additional constraints should
also be appropriate for the investigation of ``worst-case'' scenarios
and for studying possible collider signatures. 
Making use of all available information, on the other hand, would be
preferable when testing whether a certain model is actually the
``correct'' theory.

{}From the above discussion it should be obvious that it is not possible
to define a single set of benchmark scenarios that will serve all
purposes. The usefulness of a particular scenario will always depend on 
which sector of the theory (e.g.\ the Higgs or the chargino/neutralino
sector) and which physics issue is investigated (exclusion limits or
``typical'' scenarios at colliders, dark matter searches, etc.).
Accordingly, a comparison of the physics potential of different
experiments on the basis of specific benchmark scenarios is necessarily
very difficult.

The need for reconsidering the issue of defining appropriate benchmarks
for SUSY searches at the next generation of colliders becomes apparent
from the fact that the exclusion bounds in the Higgs sector of the MSSM
obtained from the Higgs search at LEP rule out several of the benchmark
points used up to now for studies of SUSY phenomenology at future
colliders. Accordingly, after the termination of the LEP program several
proposals for new benchmark scenarios for SUSY searches have been made
by different groups.

The ``Snowmass Points and Slopes'' (SPS), which we will discuss in the
following, are a set of benchmark scenarios which arose from
the 2001 ``Snowmass Workshop on the Future of
Particle Physics'' as a consensus based on different proposals recently
made 
by various groups. The SPS consist of model
lines (``slopes''), i.e.\ continuous sets of parameters depending on one 
dimensionful parameter (see below) and specific benchmark points, where
each model line goes through one of the benchmark points. The SPS should
be regarded as a recommendation for future studies of SUSY phenomenology, 
but of course are not meant as an exclusive and for all purposes sufficient
collection of SUSY models. They mainly focus on ``typical'' scenarios
within the three currently most prominent SUSY-breaking mechanisms,
i.e.\ mSUGRA, GMSB and AMSB. Furthermore they contain examples of 
``more extreme'' scenarios, e.g.\ a ``focus point''
scenario~\cite{focuspoint} with a rather heavy SUSY spectrum, indicating
in this way different possibilities for SUSY phenomenology that can 
be realized within the most commonly used SUSY breaking scenarios.

\section{Recent proposals for SUSY benchmarks}
\label{sec:two}

Before discussing the SPS in detail, we first briefly review some recent
proposals for SUSY benchmark scenarios. In \citere{BDEGMOPW}, henceforth
denoted as BDEGMOPW, a set of 13 parameter points in the CMSSM (i.e.\
the mSUGRA) scenario has been proposed according to the constraints 
arising from demanding that the lightest supersymmetric particle (LSP)
should give rise to a cosmologically acceptable dark matter relic
abundance: five points were chosen in the ``bulk'' of the cosmological
region, four points along the ``coannihilation tail'' (where a rapid
coannihilation takes place between the LSP and the (almost mass degenerate) 
next-to-lightest SUSY particle (NSLP), which is usually the
lighter $\tilde\tau$), two points were chosen in 
rapid-annihilation ``funnels'' (where an
increased annihilation cross section of the LSP results from poles due 
to the heavier neutral MSSM Higgs bosons $H$ and $A$), and two points in the
``focus-point'' region (where the annihilation cross section of the LSP
is enhanced due to a sizable higgsino component). The BDEGMOPW points
are all taken for the value of the trilinear coupling $A_0 = 0$, i.e.\ the
parameters that are varied are $m_0$, $m_{1/2}$, $\tan\beta$ and
$\mathrm{sign}(\mu)$. They were in particular chosen to span a wide
range of $\tan\beta$ values.

The constraints from the LEP Higgs search and the measurement of 
$b \to s \gamma$ have been imposed
for all of the BDEGMOPW points, while the $g_{\mu}-2$ constraint was 
not enforced (at the time of the proposal of the BDEGMOPW points only
the points in the ``bulk'' of the cosmological region were in agreement
with the $g_{\mu}-2$ constraint, while taking into account the updated SM value
for $g_{\mu}-2$~\cite{gminus2th} all but one of the BDEGMOPW points
satisfy the $g_{\mu}-2$ constraint at the $2 \sigma$ level). The
``bulk'' of the cosmological region and the low-mass portion of the
``focus point'' region are favored if fine-tuning constraints are applied.

The ``Points d'Aix'' is a different set of benchmark points, which were 
proposed in the framework of the Euro-GDR SUSY Workshop~\cite{aix}.
It consists of eleven benchmark points, out of which six belong to the mSUGRA
scenario, four to the GMSB scenario and one to the AMSB scenario. The
constraints from the LEP Higgs search and the electroweak precision data     
have been applied to all benchmark points. For the mSUGRA points
further constraints from $b \to s \gamma$, $g_{\mu}-2$, and
cosmology have been used, while for the GMSB points the 
constraints from $b \to s \gamma$ and $g_{\mu}-2$ have been taken into
account. No further constraints have been applied for the AMSB point.

In \citere{modelline} a set of eight ``model lines'' in the mSUGRA,
GMSB and AMSB scenarios has been proposed. 
The model lines were designed for studying typical SUSY signatures as a
function of the SUSY scale. Accordingly, each model line depends on
one dimensionful parameter, which sets the overall SUSY scale, while
$\tan\beta$ and $\mathrm{sign}(\mu)$ are kept fixed for each model line.
The other dimensionful parameters in each SUSY-breaking scenario are
taken to scale linearly with the parameter being varied along the model
line. 
Since the main focus in this approach lies in investigating
typical SUSY signatures, neither constraints from Higgs and SUSY
particle searches nor from $b \to s \gamma$, $g_{\mu}-2$, or cosmology
were applied. Four of the model lines refer to the mSUGRA scenario, one
corresponds to an mSUGRA-like scenario with non-unified gaugino masses,
two model lines are realizations of the GMSB scenario, and one of the
AMSB scenario.


\section{The Snowmass Points and Slopes (SPS)}

The Snowmass Points and Slopes (SPS) are based on an attempt to merge
the features of the above proposals for different benchmark
scenarios into a subset of commonly accepted benchmark scenarios. They
consist of benchmark points and model lines (``slopes''). There are 
ten benchmark points, from which six correspond to an mSUGRA
scenario, one is an mSUGRA-like scenario with non-unified gaugino
masses, two refer to the GMSB scenario, and one to the AMSB scenario. 
Seven of these benchmark points are attached to model lines, while the
remaining three are supplied as isolated points (one could of course
also define model lines going through these points, but since studying a
model line will require more effort than studying a single point, it
seemed unnecessary to equip every chosen benchmark point with a model
line). In studying the benchmark scenarios the model lines should prove
useful in performing more general analyses of typical SUSY signatures, 
while the specific points indicated on the lines are proposed to be 
chosen as the first sample points for very detailed (and thus 
time-consuming) analyses. The concept of a model line means of course
that more than just one point should be studied on each line. 
Results along the model lines can often then be roughly
estimated by interpolation.

An important aspect in the philosophy behind the benchmark scenarios is 
that the low-energy MSSM parameters 
should be regarded as the actual
benchmark rather than the high-energy input parameters $m_0$, $m_{1/2}$,
etc. Thus, specifying the benchmark scenarios in terms of the latter
parameters is merely understood as an abbreviation for the low-energy
phenomenology.

The relevant low-energy parameters are the soft SUSY-breaking parameters
in the diagonal entries of the sfermion mass matrices (using the
notation of the first generation),
\begin{equation}
M_{\tilde q1_L}, M_{\tilde d_R}, M_{\tilde u_R}, M_{\tilde e_L},
M_{\tilde e_R},
\label{eq:param1}
\end{equation}
and analogously for the other two generations, as well as
\begin{equation}    
A_t, A_b, A_{\tau}, \ldots , M_1, M_2, M_{\tilde g}, \mu, M_A,
\tan\beta,
\label{eq:param2}
\end{equation}
where the $A_i$ are the trilinear couplings, $M_1$, $M_2$ are the
electroweak gaugino mass parameters, $M_{\tilde g}$ is the
gluino mass, and $M_A$ is the mass of the $\cp$-odd neutral Higgs boson.

Our convention for the sign of $\mu$ is such that the neutralino and
chargino mass matrices have the following form
\beq
{\bf M}_{\widetilde \chi^0} =
\pmatrix{M_1 & 0 & -g'v_d/\sqrt{2}& g'v_u/\sqrt{2} \cr
0 & M_2 & gv_d/\sqrt{2} & - gv_u /\sqrt{2} \cr
-g'v_d/\sqrt{2} & gv_d/\sqrt{2} & 0 & -\mu \cr
g'v_u/\sqrt{2} & -gv_u/\sqrt{2} & -\mu & 0 },
\qquad
{\bf M}_{\widetilde \chi^\pm} = \pmatrix{M_2 & g v_u\cr
                         g v_d & \mu} .
\eeq

In order to relate the high-energy input parameters to the
corresponding low-energy MSSM parameters 
specified in \refeqs{eq:param1}, (\ref{eq:param2}),
a certain standard has to be chosen.
It was agreed that this standard should be version 7.58 of the program
{\sl ISAJET}~\cite{isajet}. It should be stressed at this point that
the definition of this standard contains a certain degree of
arbitrariness. In particular, for the purpose of defining certain
spectra as benchmarks, the issue of how accurately high-energy input
parameters can be related (via renormalization group running) to the 
corresponding low-energy parameters in different programs
(e.g.\ {\sl ISAJET}, 
{\sl SUSYGEN}~\cite{susygen}, {\sl SUSPECT}~\cite{suspect}, 
{\sl SOFTSUSY}~\cite{softsusy}, {\sl SUITY}~\cite{suity},
{\sl BMPZ}~\cite{bmpz}, etc.) is of
minor importance and therefore has not been addressed in the context of
the SPS.
Once a standard has been defined for relating the high-energy 
input parameters to the low-energy MSSM parameters, the way the latter 
were obtained and the precise values of the high-energy input parameters 
are no longer relevant. 

In order to perform the analysis of the SPS benchmark scenarios with
a program like {\sl PYTHIA}~\cite{pythia} or {\sl HERWIG}~\cite{herwig}, 
it is the easiest to
use the output of {\sl ISAJET 7.58} for the parameters specified
in \refeqs{eq:param1}, (\ref{eq:param2}) directly as input for these
programs. Alternatively, if one prefers to use the high-energy
parameters $m_0$, $m_{1/2}$, etc.\ as input in a program like 
{\sl SUSYGEN}, one should make sure that the low-energy parameters of 
\refeq{eq:param1}, (\ref{eq:param2}) agree within reasonable precision
with the actual benchmark values. If using the input values $m_0$, $m_{1/2}$,
etc.\ given below in a different program 
leads to a significant deviation in the parameters of
\refeqs{eq:param1}, (\ref{eq:param2}), these high-energy input
parameters should be adapted such that the low-energy 
parameters are brought into approximate agreement.
Since the low-energy MSSM parameters corresponding to {\sl ISAJET 7.58}
have been frozen as benchmarks by definition, an appropriate adaptation 
will also be necessary for upgrades of {\sl ISAJET} beyond version 7.58.

While it appears to be reasonable to fix certain sets of low-energy
MSSM parameters as benchmarks by definition (which in principle could have 
been done without resorting at all to scenarios like mSUGRA, GMSB and
AMSB), it on the other hand doesn't seem justified to freeze the
particle spectra, branching ratios, etc.\ obtained from these low-energy
MSSM parameters as well. It is obvious that no single program exists
which represents the current ``state of the art'' for computing all 
particle masses and branching ratios, and it should of course also be
possible to take future improvements into account. The 
level of accuracy of the theoretical predictions presently 
implemented in a
multi-purpose program like {\sl ISAJET} will not always be sufficient.
This refers in particular to the MSSM Higgs sector, where it will
usually be preferable to resort to dedicated programs like
\fh~\cite{feynhiggs}, \subh~\cite{subh}, or {\sl HDECAY}~\cite{hdecay}
for cross-checking.

For the evaluation of the mass spectra and decay branching ratios from
the MSSM benchmark parameters one should therefore choose an appropriate
program according to the specific requirements of the analysis that is
being performed. If detailed comparisons between different experiments or
different colliders are carried out, it would clearly be advantageous to
use the same results for the mass spectra and the branching ratios.

Concerning the compatibility with external constraints, all benchmark
points corresponding to the mSUGRA scenario give rise to a
cosmologically acceptable dark matter relic abundance (according to the 
bounds applied in \citeres{BDEGMOPW,aix}, i.e.\
$0.1 \leq \Omega_{\chi} h^2 \leq 0.3$ for the BDEGMOPW points and
$0.025 < \Omega_{\chi} h^2 < 0.5$ for the ``Points d'Aix''). 
In all SPS scenarios $\mu > 0$ 
has been chosen. Within mSUGRA models, positive values of $\mu$ lead to
values of $b \to s \gamma$ and $g_{\mu}-2$ which, within our present
theoretical understanding, are consistent with the current experimental
values of these quantities over a wide parameter range. While there is 
in general a slight preference for $\mu > 0$, one certainly cannot regard 
the case $\mu < 0$ as being experimentally excluded at present. We have
nevertheless restricted to scenarios with positive $\mu$, since choosing
$\mu$ negative does not lead to new characteristic experimental
signatures as compared to the case with $\mu > 0$. 

Taking the updated SM value for $g_{\mu}-2$~\cite{gminus2th} into
account, the allowed 2-$\sigma$ range for SUSY contributions to 
$a_{\mu} \equiv (g_{\mu}-2)/2$ is currently $-6 \times 10^{-10} <
a_{\mu} < 58 \times 10^{-10}$. Accordingly, at present no upper bound 
on the SUSY masses can be inferred from the $g_{\mu}-2$ constraint, but 
only a rather mild lower bound. For the constraint from 
$b \to s \gamma$, the bound $2.33 \times 10^{-4} < \mbox{BR}(b \to s
\gamma) < 4.15 \times 10^{-4}$ has been used for the BDEGMOPW
mSUGRA points~\cite{BDEGMOPW}, while $2 \times 10^{-4} < \mbox{BR}(b \to
s \gamma) < 5 \times 10^{-4}$ has been used for the mSUGRA and GMSB
points of the ``Points d'Aix''~\cite{aix}.

The main qualitative difference between the SPS (and also the recent
proposals for post-LEP benchmarks in \citeres{modelline,BDEGMOPW,aix}) 
and the benchmarks used so far for
investigating SUSY searches at the LHC, the Tevatron and a future Linear
Collider is that scenarios with small
values of $\tan\be$, i.e.\ $\tan\be \lsim 3$, are disfavored as a result
of the Higgs exclusion bounds obtained at LEP. Consequently, there is
more focus now on scenarios with larger values of $\tan\be$ than in
previous studies. Concerning the SUSY phenomenology, intermediate and
large values of $\tan\be$, $\tan\be \gsim 5$,
have the important consequence that there is
in general a non-negligible mixing between the two staus (and an even
more pronounced mixing in the sbottom sector), leading to a significant mass
splitting between the two staus so that the lighter stau becomes the
lightest slepton. Neutralinos and charginos therefore decay
predominantly into staus and taus, which is experimentally more
challenging than the dilepton signal resulting for instance from the decay 
of the second lightest neutralino into the lightest neutralino and a
pair of leptons of the first or the second generation.

Large values of $\tan\be$ can furthermore have important consequences
for the phenomenology in the Higgs sector, as the couplings of the heavy
Higgs bosons $H$, $A$ to down-type fermions are in general enhanced. 
For sizable values of $\mu$ and $m_{\tilde g}$ the $hb\bar b$ coupling
receives large radiative corrections from gluino loop corrections, 
which in particular affect the branching ratio BR($h \to \tau^+\tau^-$).

In the following we list the SPS benchmark scenarios. The value of the
top-quark mass in all cases is chosen to be $\mt = 175$~GeV.

\begin{figure}[t]
\mbox{} \hspace{.7cm} SPS 1a \hspace{7.3cm} SPS 1b
\includegraphics[height=7.5cm,width=8cm]{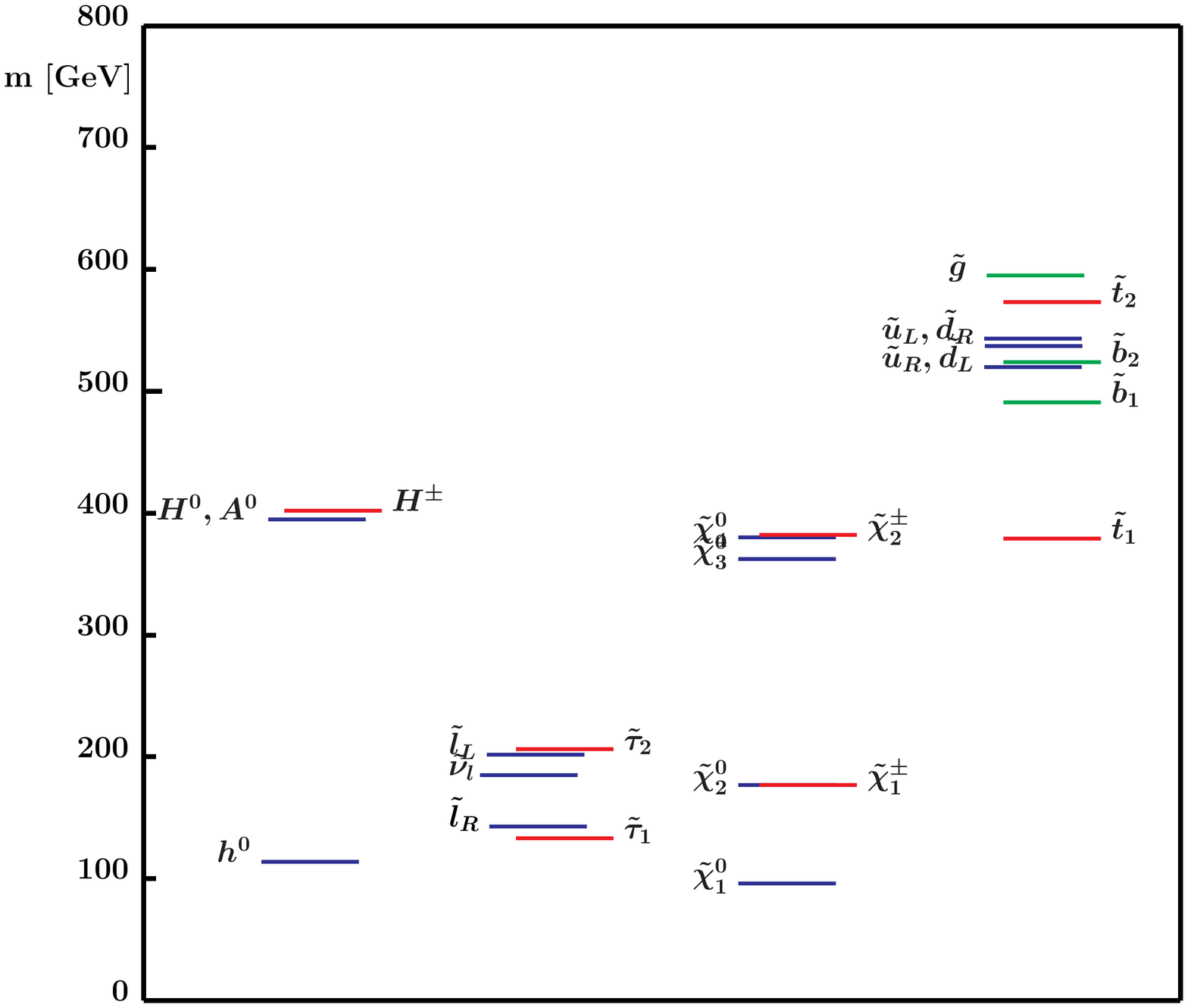} \hspace{.5em}
\includegraphics[height=7.5cm,width=8cm]{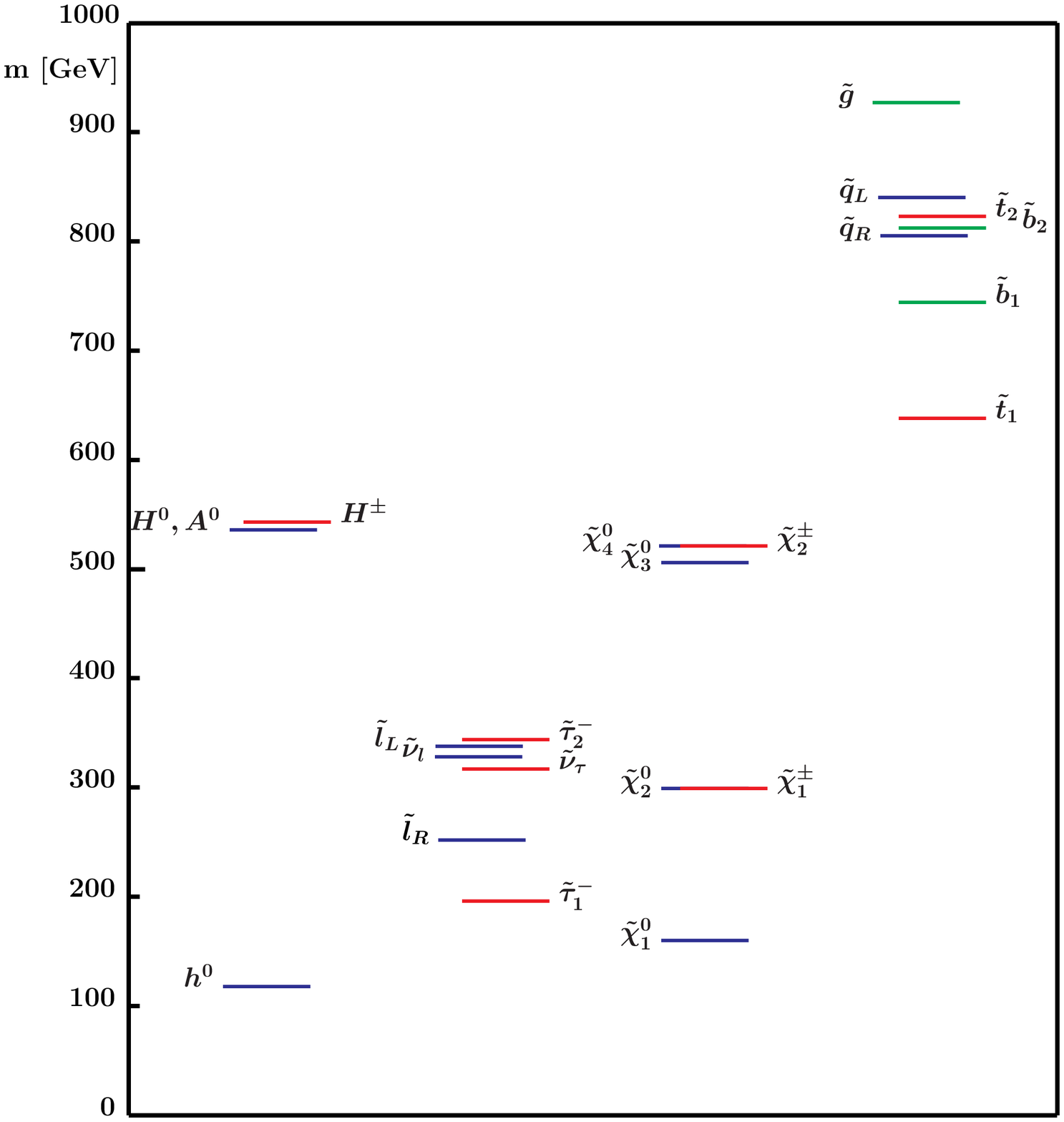}\\[1.5em]
\hspace{.7cm} SPS 2 \hspace{7.3cm} SPS 3\\
\includegraphics[height=7.5cm,width=8cm]{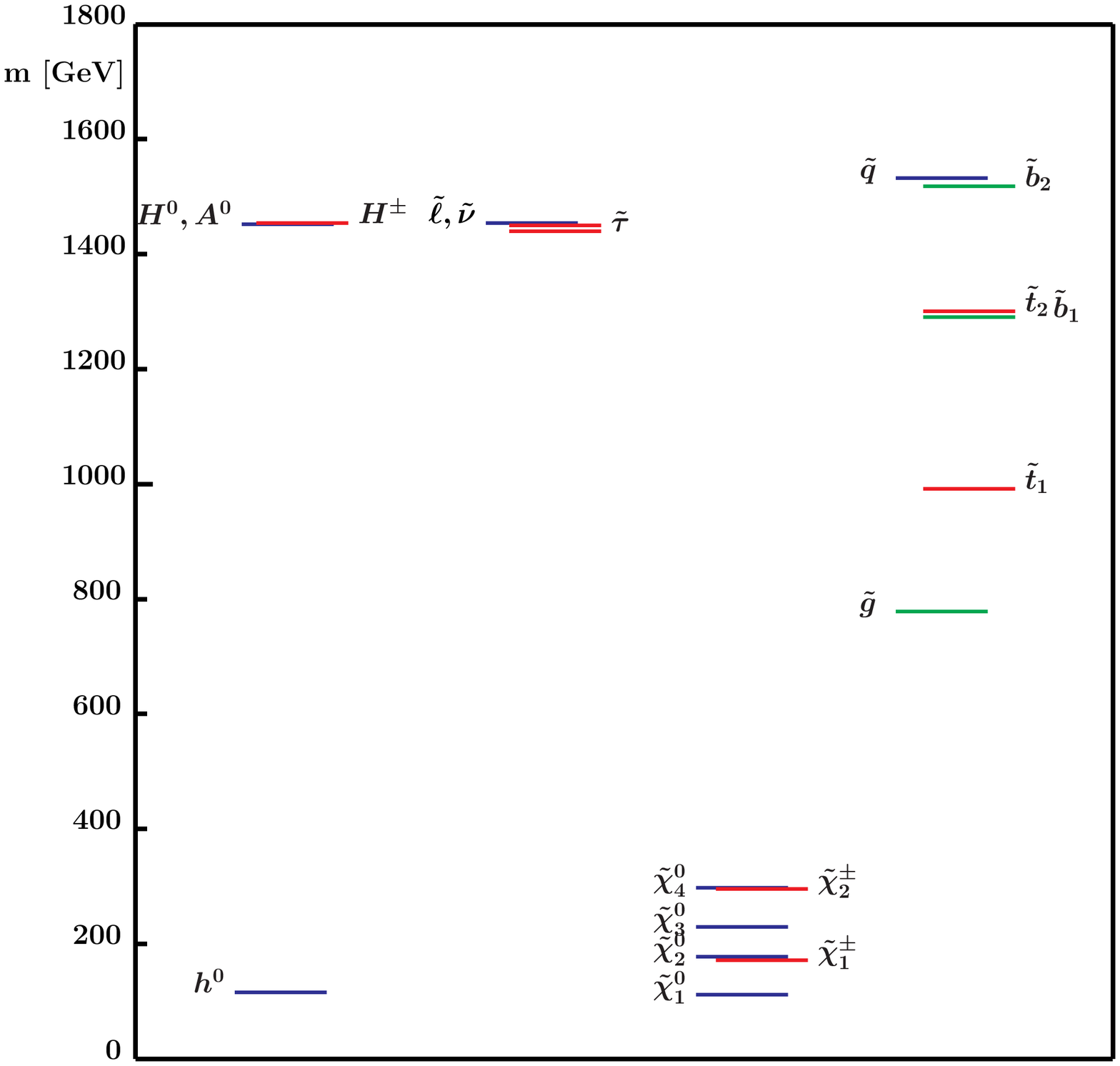} \hspace{.5em} 
\includegraphics[height=7.5cm,width=8cm]{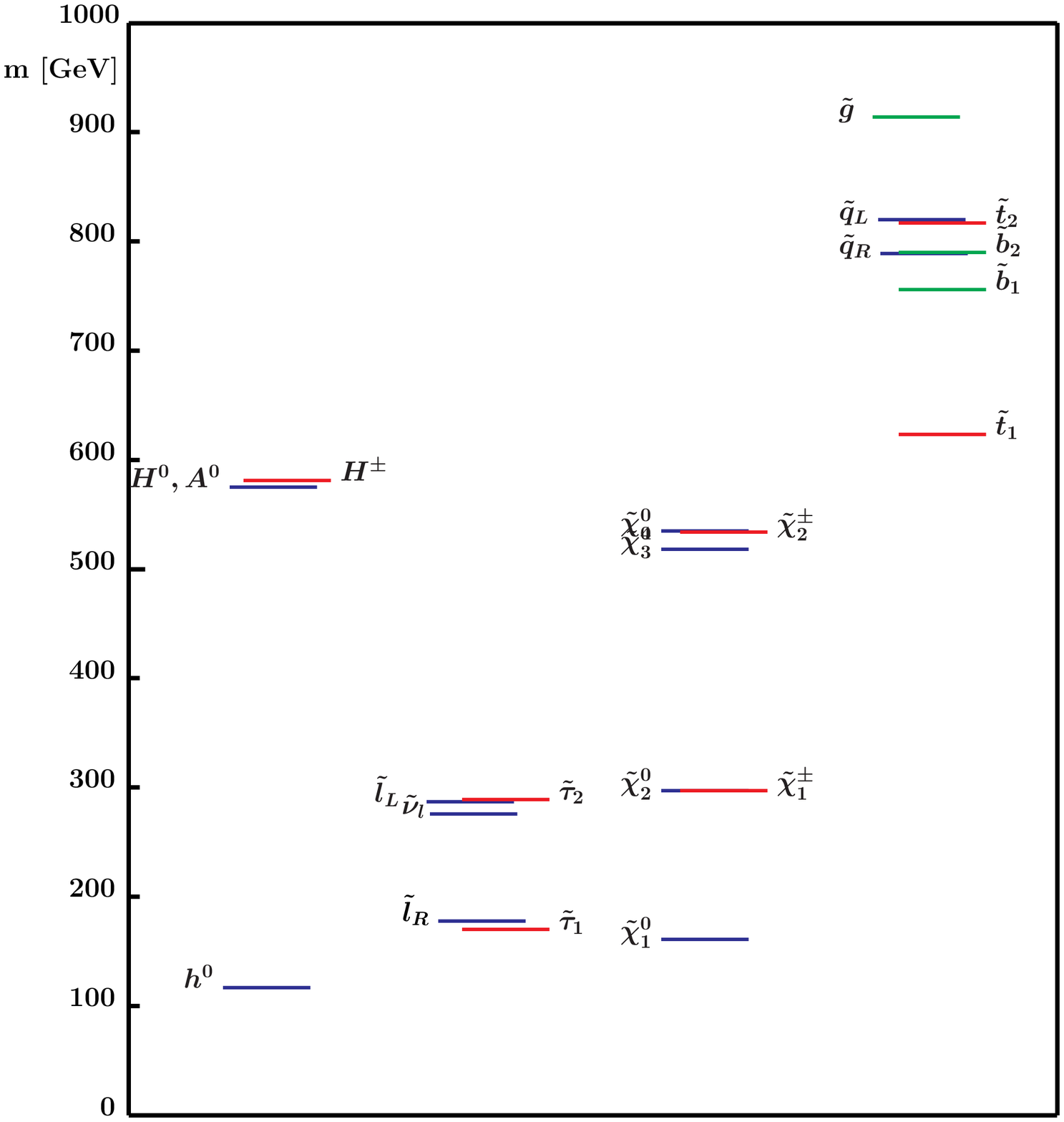} 
\caption{The SUSY particle spectra for the benchmark points
corresponding to SPS 1a, SPS 1b, SPS 2 and SPS 3 as obtained with
{\sl ISAJET 7.58} (see \citere{ulinabil}).
}
\label{fig1}
\end{figure}

\begin{figure}[t]
\mbox{} \hspace{.7cm} SPS 4 \hspace{7.3cm} SPS 5
\includegraphics[height=7.5cm,width=8cm]{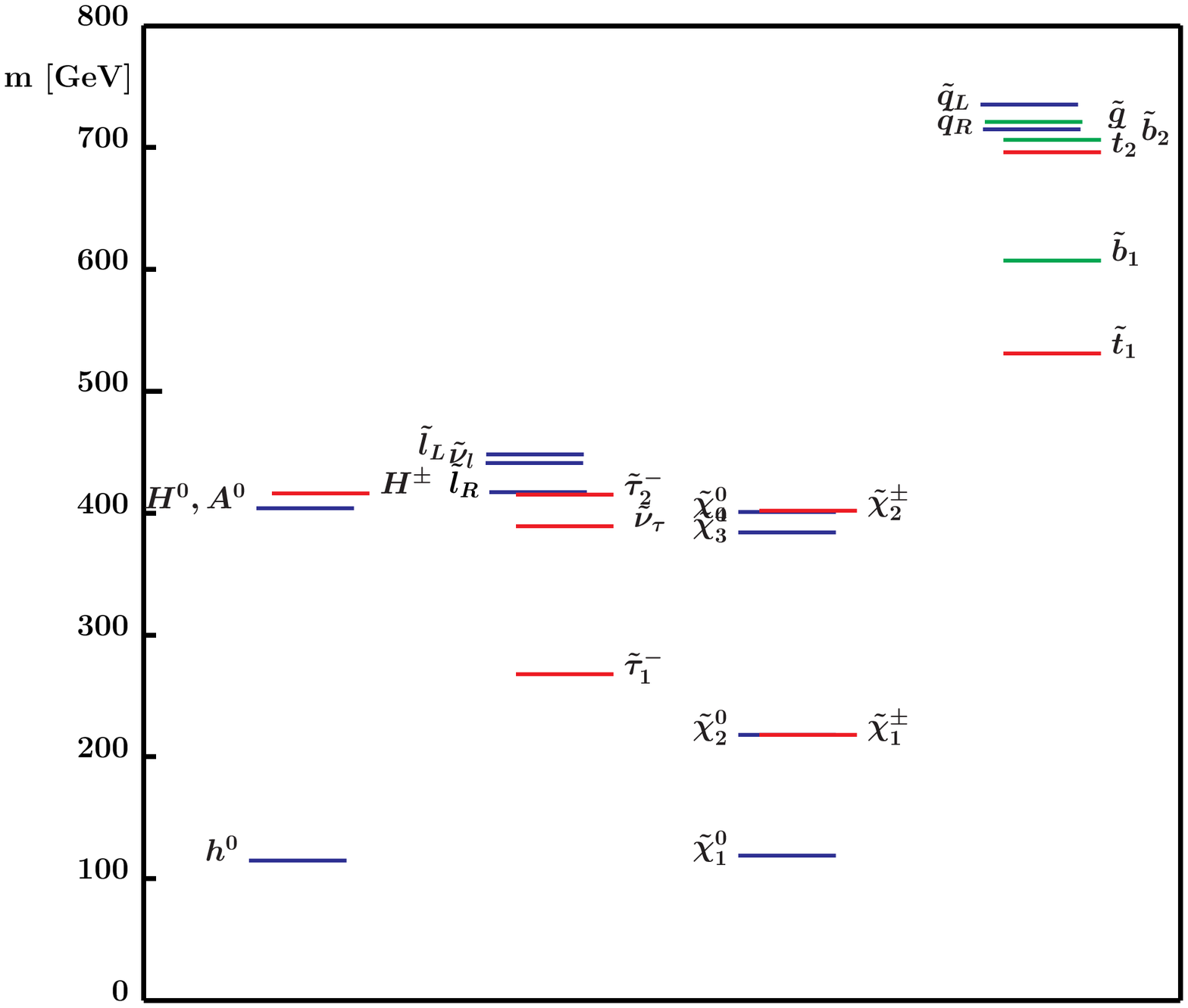} \hspace{.5em} 
\includegraphics[height=7.5cm,width=8cm]{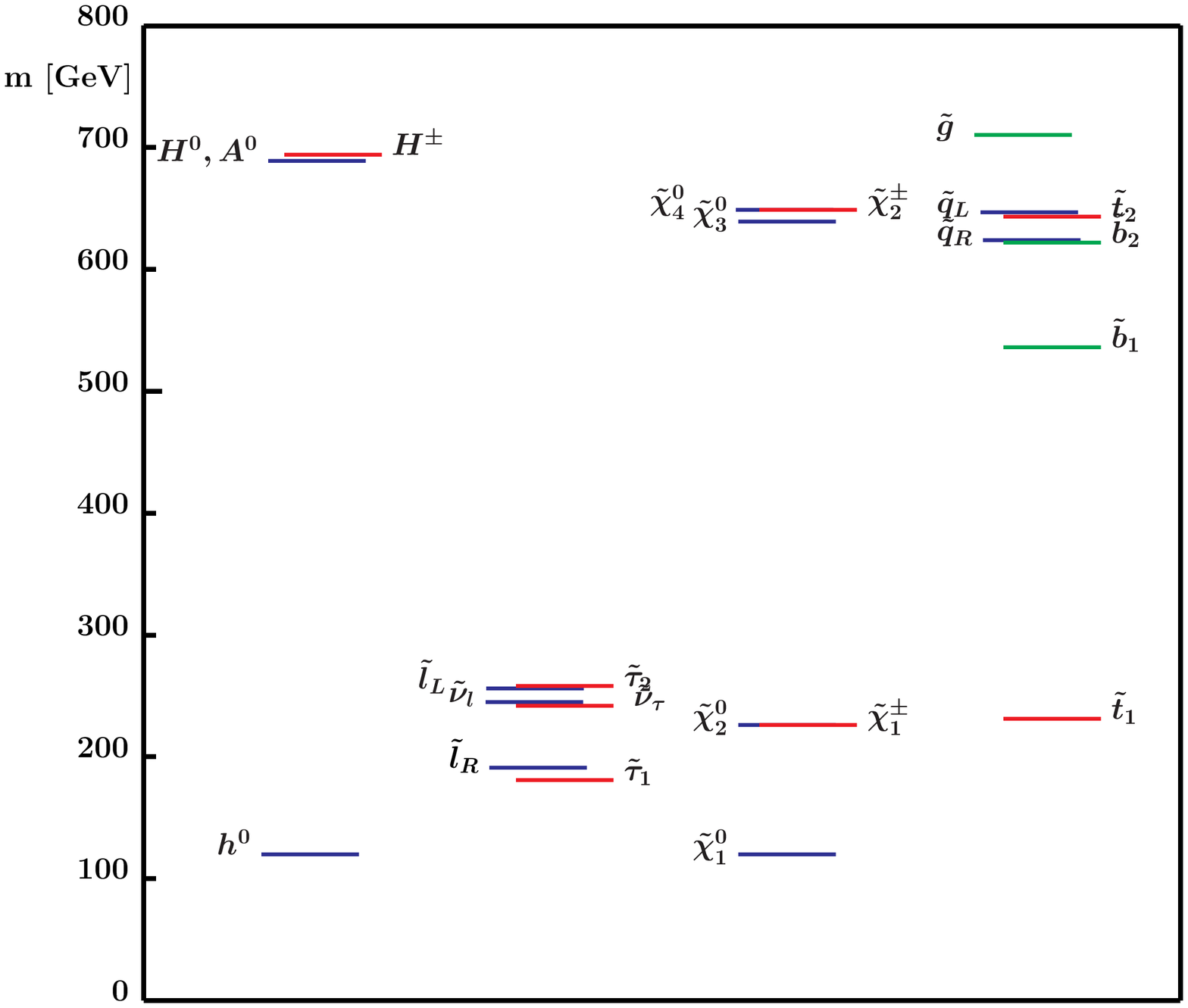} \\[1.5em]
\hspace{.7cm} SPS 6 \hspace{7.3cm} SPS 7\\
\includegraphics[height=7.5cm,width=8cm]{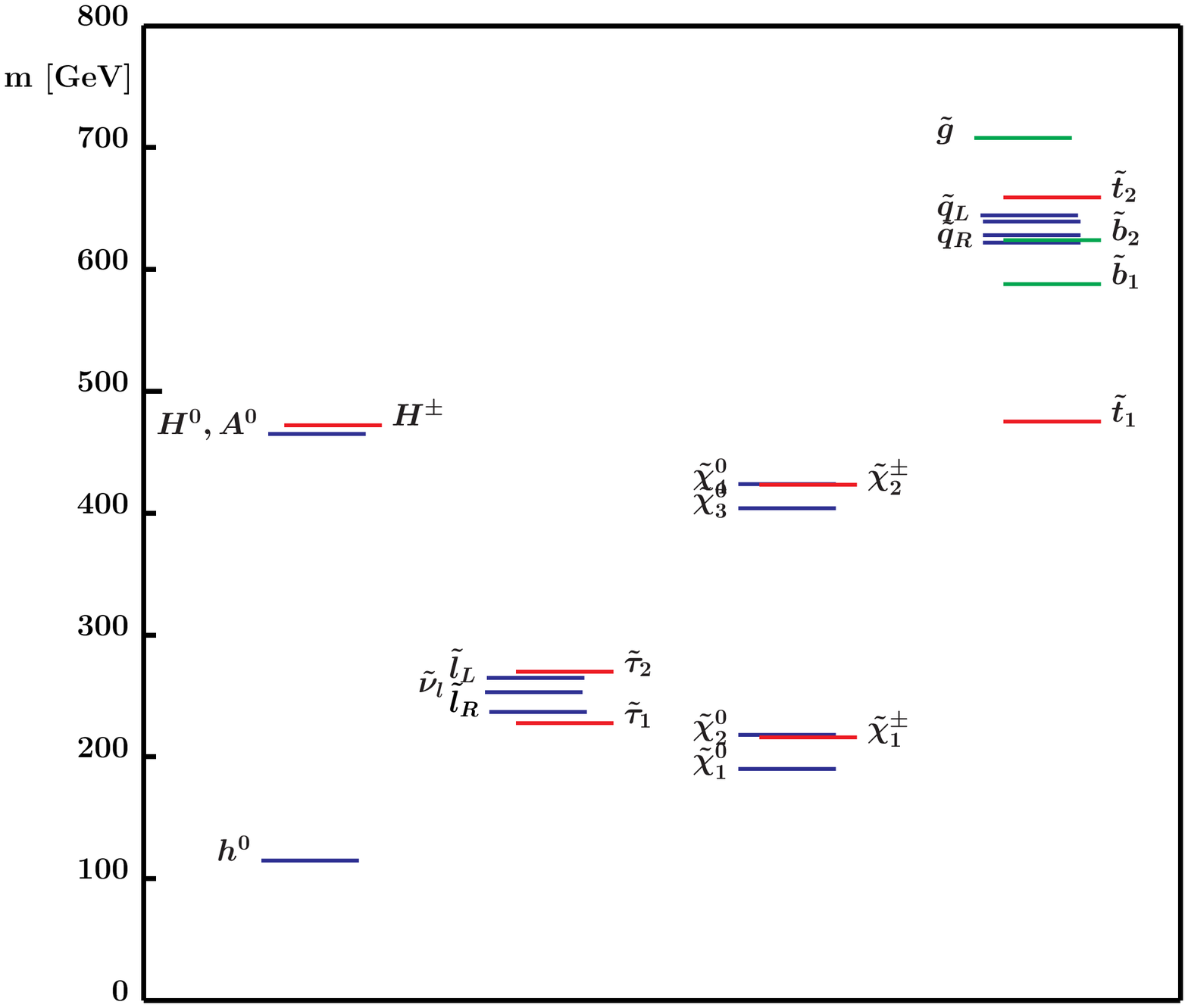} \hspace{.5em}
\includegraphics[height=7.5cm,width=8cm]{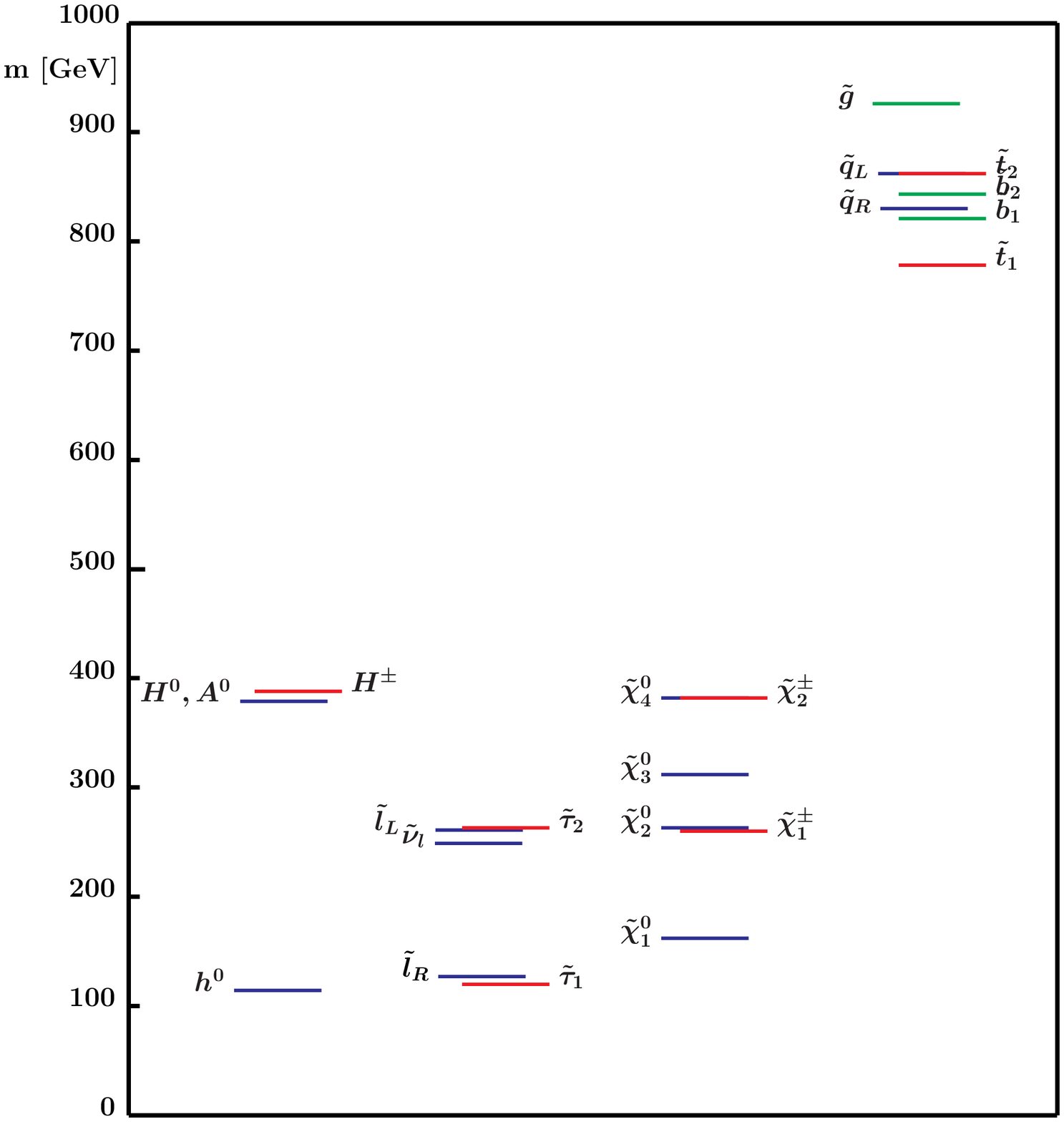}
\caption{The SUSY particle spectra for the benchmark points
corresponding to SPS 4, SPS 5, SPS 6 and SPS 7 as obtained with
{\sl ISAJET 7.58} (see \citere{ulinabil}).
}
\label{fig2}
\end{figure}

\begin{figure}[t]
\mbox{} \hspace{.7cm} SPS 8 \hspace{7.3cm} SPS 9
\includegraphics[height=7.5cm,width=8cm]{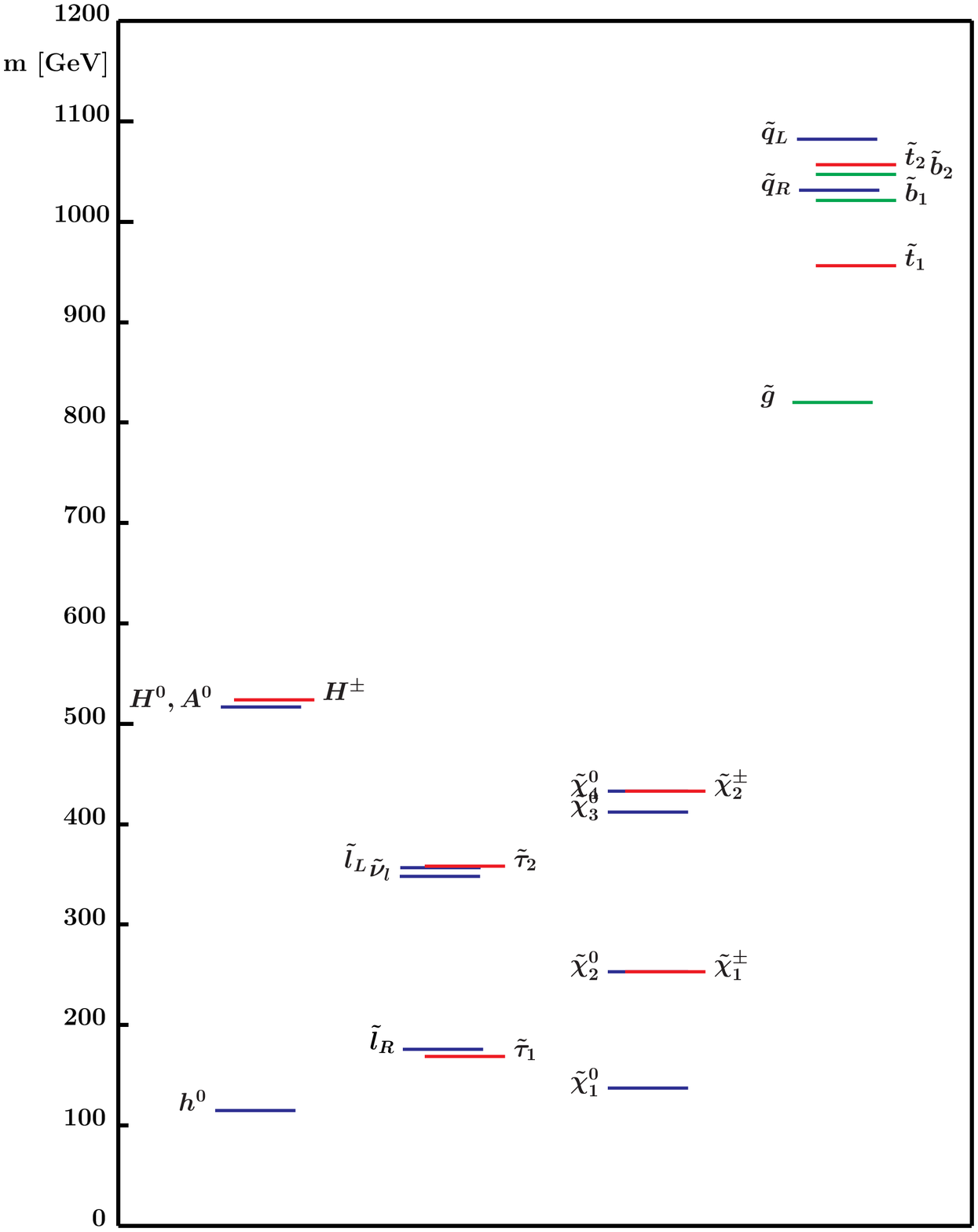} \hspace{.5em} 
\includegraphics[height=7.5cm,width=8cm]{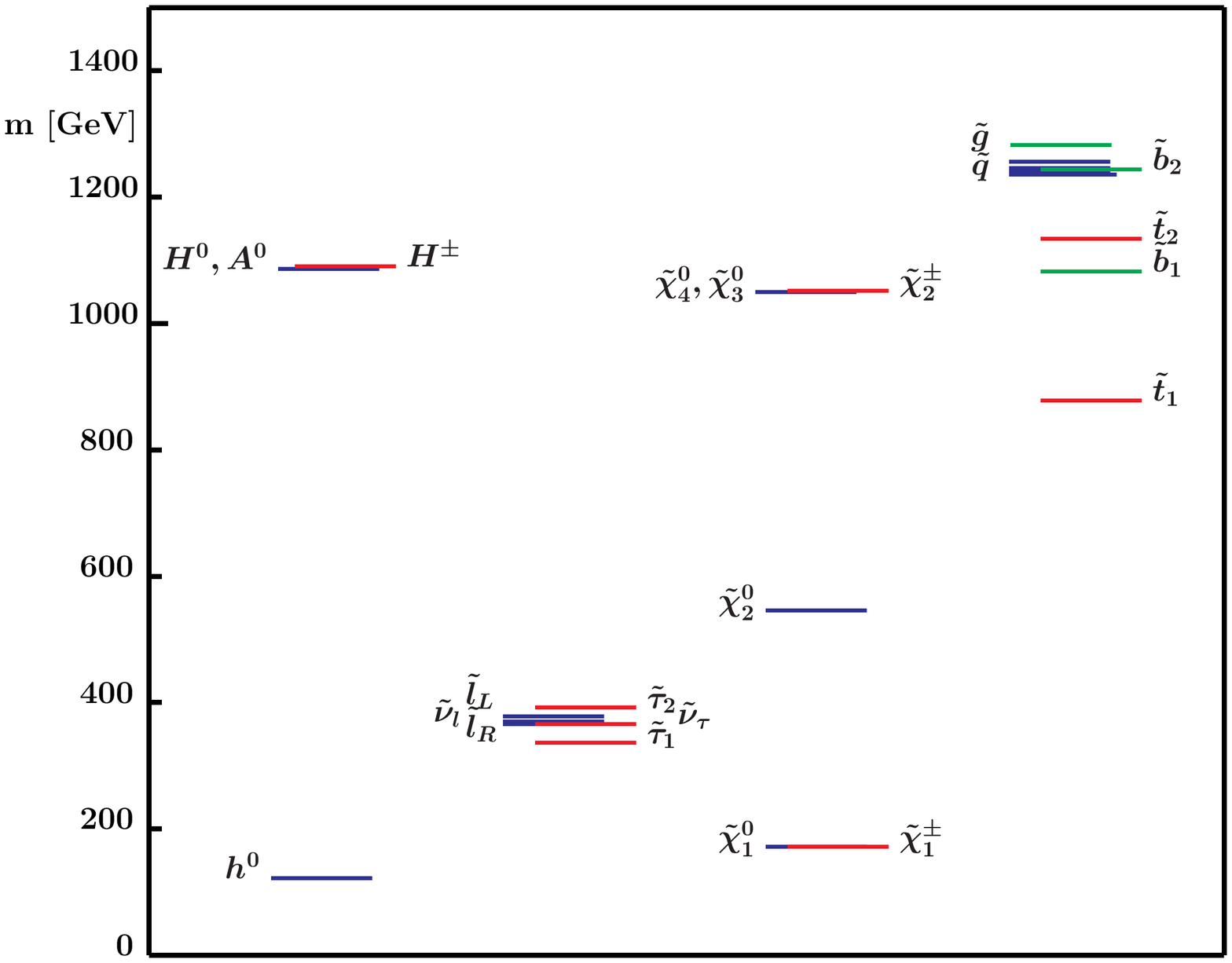}
\caption{The SUSY particle spectra for the benchmark points
corresponding to SPS 8 and SPS 9 as obtained with
{\sl ISAJET 7.58} (see \citere{ulinabil}).
}
\label{fig3}
\end{figure}

\begin{description}

\item[SPS 1: ``typical'' mSUGRA scenario] 

\mbox{}

This scenario consists of a ``typical'' mSUGRA point with an
intermediate value of 
$\tan\beta$ and a model line attached to it (SPS 1a) and of a ``typical'' 
mSUGRA point with relatively high $\tan\beta$ (SPS 1b). The two-points
lie in the ``bulk'' of the cosmological region. For the collider
phenomenology in particular the $\tau$-rich neutralino and chargino
decays are important.

{\bf SPS 1a:}\\[.5em]
\underline{Point:} 
$$
m_0 = 100 \GeV, \quad m_{1/2} = 250 \GeV, \quad A_0 = -100 \GeV, 
\quad \tan\beta = 10, \quad \mu > 0 .
$$
\underline{Slope:}
$$
m_0 = -A_0 = 0.4 \, m_{1/2}, \quad m_{1/2} \mbox{ varies} .
$$
The point is similar to BDEGMOPW point B. 
The slope equals model line A~\cite{modelline}.\\[1em]
{\bf  SPS 1b:}\\[.5em]
\underline{Point:}
$$
m_0 = 200 \GeV, \quad m_{1/2} = 400 \GeV, \quad A_0 = 0, \quad \tan\beta = 30,
\quad \mu > 0 .
$$
This point is the mSUGRA point 6 of the ``Points d'Aix''.\\ 

\item[SPS 2: ``focus point'' scenario in mSUGRA]

\mbox{}       

The benchmark point chosen for SPS~2 lies in the ``focus point'' region,
where a too large relic abundance is avoided by an enhanced annihilation
cross section of the LSP due to a sizable higgsino component.
This scenario features relatively heavy squarks and sleptons, while the
charginos and the neutralinos are fairly light and the gluino is lighter
than the squarks.

\underline{Point:}     
$$
m_0 = 1450 \GeV, \quad m_{1/2} = 300 \GeV, \quad A_0 = 0, \quad \tan\beta = 10,
\quad \mu > 0 .
$$
\underline{Slope:}
$$
m_0 = 2 \, m_{1/2} + 850 \GeV, \quad m_{1/2} \mbox{ varies} .
$$
The point equals BDEGMOPW point E 
and is similar to mSUGRA point 2 of the ``Points d'Aix''.
The slope equals model line F.\\

\item[SPS 3: model line into ``coannihilation region'' in mSUGRA]

\mbox{}

The model line of this scenario is directed into the ``coannihilation
region'', where a sufficiently low relic abundance can arise from
a rapid coannihilation between the LSP and the (almost mass degenerate)
NSLP, which is usually the lighter $\tilde\tau$. Accordingly, an
important feature in the collider phenomenology of this scenario is the
very small slepton--neutralino mass difference.

\underline{Point:}
$$
m_0 = 90 \GeV, \quad m_{1/2} = 400 \GeV, \quad A_0 = 0, 
\quad \tan\beta = 10, \quad \mu > 0 .
$$
\underline{Slope:}
$$
m_0 = 0.25 \, m_{1/2} - 10 \GeV, \quad m_{1/2} \mbox{ varies} .
$$
The point equals BDEGMOPW point C.
The slope equals model line H.\\

\item[SPS 4: mSUGRA scenario with large $\tan\beta$]

\mbox{}

The large value of $\tan\be$ in this scenario has an important impact on
the phenomenology in the Higgs sector. The couplings of 
$A, H$ to $b\bar{b}$ and $\tau^+\tau^-$ as well as the 
$H^{\pm} t\bar{b}$ couplings are
significantly enhanced in this scenario, resulting in particular
in large associated production cross sections for the heavy Higgs
bosons.
 
\underline{Point:}
$$
m_0 = 400 \GeV, \quad m_{1/2} = 300 \GeV, \quad A_0 = 0,
\quad \tan\beta = 50, \quad \mu > 0 .
$$
This point equals mSUGRA point 3 of the ``Points d'Aix'' and is 
similar to BDEGMOPW point L.\\

\item[SPS 5: mSUGRA scenario with relatively light scalar top quark]

\mbox{}

This scenario is characterized by a large negative value of $A_0$, which
allows consistency of the relatively low value of $\tan\beta$ with
the constraints from the Higgs search at LEP, see
\citere{Djouadi:2001yk}.

\underline{Point:}
$$
m_0 = 150 \GeV, \quad m_{1/2} = 300 \GeV, \quad A_0 = -1000,   
\quad \tan\beta = 5, \quad \mu > 0 . 
$$
This point equals mSUGRA point 4 of the ``Points d'Aix''.\\

\item[SPS 6: mSUGRA-like scenario with non-unified gaugino masses] 

\mbox{}

In this scenario, the bino mass parameter $M_1$ is larger than in the
usual mSUGRA models by a factor of $1.6$. While a bino-like neutralino is 
still the LSP, the mass difference between the lightest chargino and the
lightest two neutralinos and the sleptons is significantly reduced 
compared to the typical mSUGRA case. Neutralino, chargino and slepton decays
will feature less-energetic jets and leptons as a consequence. 
 
\underline{Point:}
\begin{eqnarray}
&& \mbox{at GUT scale: } M_1 = 480 \GeV, \quad M_2 = M_3 = 300 \GeV \non \\
&&
m_0 = 150 \GeV, \quad m_{1/2} = 300 \GeV, \quad A_0 = 0,
\quad \tan\beta = 10, \quad \mu > 0 .  \non
\end{eqnarray} 
\underline{Slope:}
$$
M_3({\rm GUT}) = M_2({\rm GUT}), \quad 
M_1({\rm GUT}) = 1.6 \, M_2({\rm GUT}), \quad 
m_0 = 0.5 \, M_2({\rm GUT}), 
\quad M_2({\rm GUT}) \mbox{ varies} .
$$
The slope equals model line B.\\

\item[SPS 7: GMSB scenario with $\tilde \tau$ NLSP]

\mbox{}

The NLSP in this GMSB scenario is the lighter stau, with allowed three
body decays of right-handed selectrons and smuons into it.
The decay of the NLSP into the Gravitino and the $\tau$ in this scenario
can be chosen to be prompt, delayed or quasi-stable.
 
\underline{Point:}
$$
\Lambda = 40 \TeV, \quad M_{\rm mes} = 80 \TeV, \quad N_{\rm mes} = 3,
\quad \tan\beta = 15, \quad \mu > 0 .  
$$
\underline{Slope:}
$$
M_{\rm mes}/ \Lambda = 2, \quad \Lambda \mbox{ varies} .
$$
The point equals GMSB point 1 of the ``Points d'Aix''.
The slope equals model line D.\\

\item[SPS 8: GMSB scenario with neutralino NLSP]

\mbox{}      

The NLSP in this scenario is the lightest neutralino. The second
lightest neutralino has a significant branching ratio into $h$ when
kinematically allowed.
The decay of the NLSP into the Gravitino (and a photon or a $Z$~boson)
in this scenario can be chosen to be prompt, delayed or quasi-stable.


\underline{Point:}
$$
\Lambda = 100 \TeV, \quad M_{\rm mes} = 200 \TeV, \quad N_{\rm mes} = 1,
\quad \tan\beta = 15, \quad \mu > 0 .
$$
\underline{Slope:}
$$
M_{\rm mes}/ \Lambda = 2, \quad \Lambda \mbox{ varies} .
$$
The point equals GMSB point 2 of the ``Points d'Aix''.
The slope equals model line E.\\

\item[SPS 9: AMSB scenario]

\mbox{}      

This scenario features a very small neutralino--chargino mass
difference, which is typical for AMSB scenarios. Accordingly, the LSP is
a neutral wino and the NLSP a nearly degenerate charged wino. The NLSP
decays to the LSP and a soft pion with a macroscopic decay length, as
much as 10~cm.

\underline{Point:}
$$
m_0 = 450 \GeV, \quad m_{\rm aux} = 60 \TeV, 
\quad \tan\beta = 10, \quad \mu > 0 .
$$
\underline{Slope:}
$$
m_0 = 0.0075 \, m_{\rm aux}, \quad m_{\rm aux} \mbox{ varies} . 
$$
The slope equals model line G.

\end{description}

\begin{table}[t]
\begin{center}
\begin{tabular}{cccccccc}
\hline\hline
SPS & \multicolumn{6}{c}{Point \hspace{3em}} & Slope\\ 
\hline\hline
mSUGRA: & $m_0$ & $m_{1/2}$ & $A_0$ & $\tan\beta$ & & & \\
\hline
1a & 100  & 250 &   -100 & 10 & & & $m_0 = -A_0 = 0.4 \, m_{1/2}$, 
                                    $\; m_{1/2}$ varies\\
1b & 200  & 400 &      0 & 30 & & & \\
2  & 1450 & 300 &      0 & 10 & & & $m_0 = 2 \, m_{1/2} + 850 \GeV$,
                                    $\; m_{1/2}$ varies\\
3  &   90 & 400 &      0 & 10 & & & $m_0 = 0.25 \, m_{1/2} - 10 \GeV$,
                                    $\; m_{1/2}$ varies \\
4  &  400 & 300 &      0 & 50 & & & \\
5  &  150 & 300 &  -1000 &  5 & & & \\
\hline\hline
mSUGRA-like: & $m_0$ & $m_{1/2}$ & $A_0$ & $\tan\beta$ & 
               $M_1$ & $M_2 = M_3$ & \\
\hline      
6  &  150 & 300 &      0 & 10 & 480 & 300 & $M_1 = 1.6 \, M_2$, 
                              $m_0 = 0.5 \, M_2$, $\; M_2$ varies \\ 
\hline\hline         
GMSB: & $\Lambda/10^3$ & $M_{\rm mes}/10^3$ & $N_{\rm mes}$ & $\tan\beta$ & & & \\ 
\hline     
7  &   40 &  80 &      3 & 15 & & & $M_{\rm mes}/ \Lambda = 2$, 
                                    $\; \Lambda$ varies \\    
8  &  100 & 200 &      1 & 15 & & & $M_{\rm mes}/ \Lambda = 2$, 
                                        $\; \Lambda$ varies \\   
\hline\hline 
AMSB: & $m_0$ & $m_{\rm aux}/10^3$ & & $\tan\beta$ & & & \\ 
\hline         
9  &  450 &  60 &        & 10 & & & $m_0 = 0.0075 \, m_{\rm aux}$,
                                    $\; m_{\rm aux}$ varies \\
\hline\hline   
\end{tabular}
\end{center}
\caption{The parameters (which refer to 
{\sl ISAJET} version 7.58) for the Snowmass Points and Slopes (SPS). 
The masses and scales are given in GeV.
All SPS are defined with $\mu > 0$. The parameters $M_1$, $M_2$, $M_3$
in SPS 6 are understood to be taken at the GUT scale. The value of the
top-quark mass for all SPS is $\mt = 175$~GeV.
\label{params}}
\end{table}

For completeness, the parameters of all benchmark scenarios have been
collected in Table~\ref{params}.
The SUSY particle spectra corresponding to the benchmark points of the 
SPS as obtained with {\sl ISAJET 7.58} are shown in 
\reffis{fig1}-\ref{fig3}. 

For a detailed listing of the low-energy MSSM parameters obtained with
{\sl ISAJET 7.58} corresponding to the benchmark points specified above
we refer to \citere{ulinabil}.$^a$
\renewcommand{\thefootnote}{\alph{footnote}}
\footnotetext[1]{The results for SPS~1b are not given in
\citere{ulinabil}.}

In \citere{ulinabil} furthermore {\sl PYTHIA} and {\sl SUSYGEN} have
been used in order to derive the low-energy MSSM parameters for the
mSUGRA benchmark points of the SPS (i.e.\ using the high-energy 
parameters specified in SPS 1a, 2, 3, 4, 5 as input). These results can
be used to adapt the high-energy input parameters in {\sl PYTHIA} and
{\sl SUSYGEN} such that the actual benchmarks are closely resembled.
For SPS 1a, 3, and 5 quite good agreement (typically within 10\%) 
between the low-energy MSSM parameters obtained with {\sl ISAJET 7.58},
{\sl PYTHIA 6.2/00} and {\sl SUSYGEN 3.00/27} has been found. For the
high-energy input parameters corresponding to SPS~2 and 4, which involve 
more extreme values (large $m_0$ in SPS~2 and large $\tan\beta$ in
SPS~4), rather drastic deviations between low-energy parameters obtained 
with the 
three programs can occur (in the chargino and neutralino sector for
SPS~2 and in the Higgs and third generation sfermion sector for SPS~4),
indicating that the theoretical uncertainties in relating the
high-energy input parameters to the low-energy MSSM parameters are very
large in these cases. 
Consequently, some adaptations of the high-energy input parameters
will be necessary when analyzing 
SPS~2 and 4 with different codes in order to match the actual
benchmarks.

In \citere{ulinabil} also the particle spectra and decay branching
ratios obtained with {\sl ISAJET 7.58}, {\sl PYTHIA 6.2/00} and {\sl
SUSYGEN 3.00/27} have been compared. For SPS 6 -- 9,
where the benchmark values of the low-energy MSSM parameters have been
used as input for {\sl PYTHIA} and {\sl SUSYGEN}, a good overall
agreement in the particle spectra and branching ratios between the three
programs has been found.
For a similar analysis, in which the outputs of different codes 
are compared for some of the model lines specified above, see
\citere{benallanach}.

As mentioned above, in order to allow detailed comparisons between 
future studies based on the SPS it is not only important that the
correct values for the actual benchmark parameters specified 
in \refeqs{eq:param1}, (\ref{eq:param2}) are used, but also the mass
spectra and branching ratios that were used in the studies should be
indicated.

\section{Conclusions}

Detailed experimental simulations in the search for supersymmetric
particles make it often necessary to restrict oneself to specific
benchmark scenarios. The usefulness of a particular benchmark scenario
depends on the physics issue being investigated, and the question of which
points or parameter lines should be selected from a multi-dimensional
parameter space is to a considerable extent a matter of taste.
After the completion of the LEP program several sets of benchmark
scenarios for SUSY searches
have been proposed as a guidance for experimental analyses at
the Tevatron, the LHC and future lepton and hadron colliders. 
These proposals have been
discussed at the ``Snowmass Workshop on the Future of Particle
Physics'', and have briefly been reviewed in this paper. 

As an outcome of the Snowmass Workshop the ``Snowmass Points and Slopes''
(SPS) have been agreed upon as an attempt to merge elements of the
different existing proposals into a common set of benchmark scenarios. 
The SPS, as spelled out in this paper, consist of a set of benchmark
points and model lines (``slopes'') within the mSUGRA, GMSB and AMSB 
scenarios, where each model line contains one of the benchmark points.
We hope that this collection of benchmark scenarios will prove useful 
in future experimental studies.

\begin{acknowledgments}
Fermilab is operated by Universities Research Association Inc.\
under contract no.\ DE-AC02-76CH03000 with the U.S.\ Department of
Energy.
This work was supported in part by the European Community's Human
Potential Programme under contract HPRN-CT-2000-00149 Physics at Colliders.
J.F.G.\ is supported, in part, by the U.S.\ Department of Energy,
Contract DE-FG03-91ER-40674, and by
the Davis Institute for High Energy Physics.
H.E.H.\ is supported in part by a grant from the U.S.\ Department of
Energy.
The work of J.L.H.\ is supported by the U.S.\ Department of Energy, Contract
DE-AC03-76SF00515.
J.K.\ is supported in part by the KBN Grant 5 P03B 119 20 (2001-2002).
The work of S.P.M.\ is supported in part by U.S.\ National Science Foundation
grant PHY-9970691.
F.M.\ is supported by the Fund for Scientific Research (Belgium).
S.Mo.\ would like to thank The Royal Society (London, UK) for
financial support in the form of a Conference Grant.
The work of K.A.O.\ was supported in part by DOE grant
DE--FG02--94ER--40823.
\end{acknowledgments}




\end{document}